\documentclass[journal=jacsat,manuscript=article]{achemso}

\usepackage[version=3]{mhchem} 

\author{Winston Lindqwister}
\affiliation[Duke University]
{Department of Civil and Environmental Engineering, Duke University, USA}
\email{winston.lindqwister@duke.edu}
\phone{+1 919 684 5867}
\author{Manolis Veveakis}
\affiliation[Duke University]
{Department of Civil and Environmental Engineering, Duke University, USA}
\author{Martin Lesueur}
\affiliation[Delft]{TU Delft, Delft, Netherlands}

\title[CRN for porous media]
  {Chemical homogenization for non-mixing reactive interfaces in porous media}

\abbreviations{IR,NMR,UV}

\begin{document}


\begin{abstract}
    Porous media, while ubiquitous across many engineering disciplines, is inherently difficult to characterize due to their innate stochasticity and heterogeneity. The key for predicting porous material behavior comes down to the structuring of its microstructure, where the linkages of microstructural properties to mesoscale effects remain as one of the key questions in unlocking understanding of this class of materials. One proposed method of linking scales comes down to using Minkowski functionals--geometric morphometers that describe the spatial and topological features of a convex space--to draw connections from microstructural form to mesoscale features. In this work, chemical equilibrium and kinetics on a microstructure surface were explored, with Minkowski functionals used as the basis for relating microstructural geometry to chemical performance. Using surface CRNs to model chemical behavior--a novel asynchronous cellular automaton-- linkages were found between the Minkowski functionals and equilibrium equilibrium constant, as well as properties related to the dynamics of the system's reaction quotient.
\end{abstract}

\section{Introduction}

While ubiquitous, from bone to rock to fuel cells, porous media represents a wide class of materials that remain elusive to fully characterize. While their properties at the microstructural level have shown to be intrinsically linked to mesoscale effects, the exact nature of this scaling has proven to be highly elusive\cite{Biot1941-bm, Biot2021-xq}. In order to link these effects, one approach has been to use geometric homogenization schemes to derive energetic linkages from microstructural form to behavior\cite{Charalambakis2010-cl, Groen2018-tn, Guevel2021-jf}.
\\
Minkowski functionals are geometric morphometers, characterizing both morphology and topology of spatial patterns, that are conceptualized from the field of statistical physics\cite{Armstrong2019-fh}. They have seen wide application in describing phenomenon from the spin of galaxies\cite{Hikage2003-mp} to the permeability of porous media\cite{Slotte2020-ek}. The use of these functionals as a descriptor for meso-scale phenomenon is supported by Hadwiger's theoerm\cite{Klain1995-nv}, which guarantees that for a polyconvex, isotropic system of dimension $D$, $D+1$ Minkowski functionals can be used to sufficiently describe the behavior of the system. In particular, Minkowski functionals have shown to have a powerful connection between the geometry and the free energy of a system, creating an important linkage between structural and energetic properties of materials\cite{Klain1995-nv}.
\\
Of particular interest for porous media is quantifying their chemical behavior\cite{Mauri1991-iw, Gatica1989-je, Valdes-Parada2010-cl, Zachara2016-mo}. Chemical activity in porous media drives both immediate behavior\cite{Xiong2021-zv, Garcia2005-uv} and long-term performance\cite{Zen1963-ka, Evans2013-xl, Adamo2000-sd}. Due to the many complexities typical of porous systems, from the many interacting chemical species to the inherent challenges of accurately modeling chemical surface interactions, homogenization schemes for succinct characterization of microstructural performance are of perticular interest\cite{Mikelic2018-vd, Kumar2016-yy, Allaire2010-ob}.
\\
In the world of modeling well-mixed systems, the classic approach to homogenizing complex chemical systems is through chemical reaction networks\cite{Feinberg2019-jo, Unsleber2020-cu}. Chemical reaction networks (CRNs) are graph-based models of dynamic chemical systems that typically organize chemical species as functions $f(x)$ and their evolution as $\dot x$ to form a continuous autonomous dynamic system of the form $\dot x = f(x)$. These models provide powerful tools in identifying reaction system steady states\cite{Tonello2018-os}, steady state stability\cite{Feinberg1988-eg}, persistence\cite{Komatsu2019-hn}, existence of stable periodic solutions\cite{Russo2006-io}, and performing model reduction\cite{Rao2013-xo, Rao2014-ag}. While these models are quite powerful in driving understanding of these complex dynamic systems, there are certain assumptions of a traditional CRN that limit their ability to fully characterize interfacial systems such a porous media. Namely, CRN models typically assume a well mixed system of comparable density across the entire domain\cite{Soloveichik2008-qa}. To address this limitation, Qian et al. proposed a novel method for implementing a CRN on a surface, applying a graph structure to a geometric boundary with CRN-like kinetics\cite{Qian2014-he}. This method, known as a surface CRN, is implemented as an asynchronous cellular automata with probabilistic transition rules that mimic a continuous-time Markov chain process. Through Qian and Winfree's work, as well as advancements from Clamons et al., surface CRNs have demonstrated the ability to form dynamic spatial patterns, operate as DNA circuits, and model adsorption and desoprtion behavior on a surface\cite{Clamons2020-yv, Qian2014-he}. Through this work, we aim to extend the implementation of these models to solid-fluid interfacial behavior on a porous microstructure.

\section{Methods}

\subsection{Surface CRNs}

A surface CRN resembles the rules of a classic CRN modeling approach, but crucially imposes spatial constraints on the manner at which reactions can occur. By definition, a surface CRN is an asynchronous, stochastic cellular automaton with CRN-like transition rules\cite{Qian2014-he}. Informally, this can be seen as a CRN where individual chemical species are localized to sites on a specific surface and may only interact with neighboring molecules\cite{Clamons2020-yv}. By a more formal definition, a surface CRN is a continuous-time Markov chain defined by a lattice $L$ of connected sites $i \in L$ with each site defined by a state $s_i$ and each site defined as $i$. The ability to switch states is determined by a set of unimolecular or bimolecular transition rules $r \in R$, where each reaction is defined as $A \rightarrow B$ or $A + B \rightarrow C + D$, with the rate of each reaction as $\lambda_r$. As an asynchronous cellular automata, each reaction occurs independently, with the ordering of these reactions processing via a queuing system. Essentially, at each frame of the simulation, the simulation grid is queried for all potential reactions that may occur based on each node's neighbors, and each potential reaction has the time for it occurring drawn from an exponential distribution. This is time to next reaction $\Delta t$ is calculated as follows:
\begin{align}
    \Delta t = \log\Bigg(\frac{1}{rand(x)}\Bigg)\Bigg(\frac{1}{\lambda_r}\Bigg)
\end{align}
with $rand(x)$ serving as the random draw from the distribution. After each time to next reaction is calculated for all candidate nodes, each node has its corresponding reaction scheduled for time $t + \Delta t$ and pushed to a priority heap queue. From here, the first reaction from the queue is popped and processed, changing the respective reactants to products. With the new map in place, the current time of the simulation is set to $t = t + \Delta t$ and all reactions in the queue involving sites changed in the aforementioned step are removed. The new site species are checked for any potential reactions and these are added to the queue as previously described, and this is repeated until a stop condition is met. Simplified, this can be seen as:
\begin{enumerate}
    \item Initialize with a global state grid at time $t = 0$.
    \item Scan each node for potential reactions that can occur, and calculate a time to next reaction $t + \Delta t$ and add it to a priority heap queue.
    \item Pop the first reacction in the queue and process reactants to products, setting the new time as $t = t + \Delta t$.
    \item Remove all reactions involving the same sites as the current reaction site in question from the queue.
    \item Scan the products in the current site for new potential reactions, and recalculate and add to the queue as in step 2.
    \item Continue from step 3 until a stop condition (such as the maximum duration of the simulation being reached, or an empty queue) has been met.
\end{enumerate}
As described in Clamons et al, the total time complexity of the simulation is $O(n + r\log w)$, where $n$ is the number of sites in the surface or the CRN, $r$ is the total number of reaction events simulated, and $w$ is the maximum number of reactions in the queue at any given time\cite{Clamons2020-yv}.
\newline
Although explicitly surface CRN reactions may only take transition rules as chemical reactions, other surface/species behavior may be emulated using the relative flexibility of what is defined as a "reaction." For example, by default, surface CRNs do not allow for diffusion of molecules. However, in this work, diffusion of mulecules is simulated using reactions of the form $X + E \xrightarrow{k} E + X$, where $X$ is the diffusing species in question, $E$ represents an exmpy site that said species can travel to, and $k$ controls the rate of diffusion.
\newline
While qualitative in nature, surface CRNs provide a simple and straightforward model of CRN-like chemistry that accounts for the geometric considerations of an interface-sensitive chemical system that a typical CRN model cannot provide. Compared to other discrete, stochastic reaction-diffusion models such as Kinetic Monte Carlo (KMC) and stochastic reaction-diffusion systems, surface CRNs come with a host of advantages and trade-offs. The primary difference between surface CRNs and other models is the requirement for species to exist in discrete spaces compared to continuous positions of species\cite{Fange}. This allows surface CRNs to naturally capture macromolecular crowding behavior, as well as to preserve local geometry of chemical reactions \cite{Sieradzki2013-un}. The relative simplicity of calculating surface CRN switching rules also make them highly parallizable--every reaction occurs in a queue and is processed one-at-a-time. One could easily segment a space into multiple surface CRNs, allowing for rapid parallel processing of large-system behavior.
\newline
For this study, a benchmark dissolution reaction was studied to understand the linkage between Minkowski functionals and chemical behavior. The benchmark dissolution reaction is of the form:
\begin{align} \label{Eq:Dissrxn}
    A + Q \rightleftharpoons 2R
\end{align}
With $A$ defined as a reactive solid species, $Q$ as a reactive liquid, and $R$ as a reaction liquid product. This reaction is seen here as a generic form of fluid-release reactions where no solids are produced and theliquid products are not mixing with preexisting fluids. This makes the transition rules of the reaction at a solid-liquid interface straightforward since no solid is retained. Indeed, this is reflected in Table \ref{tab:genericrxninput} which lists the input transition rules for the surface CRN simulator. It is to be noted that this choice of interface reaction is constraining the conclusions of the present study to non-mixing fluid-release reactions rather than to any generic interfacial reaction, which should be the subject of future works. 

\begin{table}[]
\begin{tabular}{|c|c|}
\hline
Transition Rule           & Reaction Rate \\ \hline
$A + Q \rightarrow R + R$ & 0.4           \\
$R + R \rightarrow A + Q$ & 0.1           \\
$R + Q \rightarrow Q + R$ & 1.0           \\ \hline
\end{tabular}
\caption{Transition rules for benchmark diffusion reaction.}
\label{tab:genericrxninput}
\end{table}

\subsection{Whittaker-Eilers Filter}
\begin{figure}[ht!]
    \centering
    \includegraphics[scale=0.5]{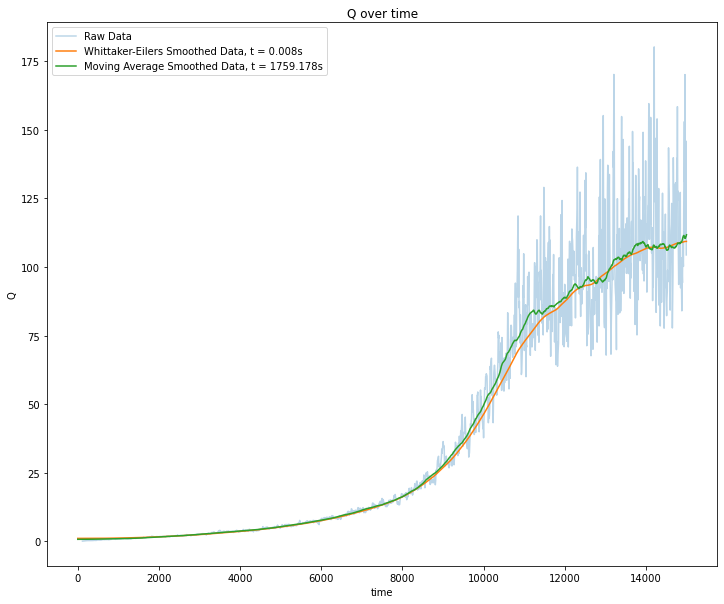}
    \caption{Comparison of filter results on a sample $Q$ calculation over 14.7 million datapoints, comparing a moving average series calculated with a window size of 1,000,000 and a Whittaker-Eilers series with $\lambda = 80,000$. Note the time to completion of the smoothing algorithms, with the Whittaker-Eilers smoothing function calculating at a speed $10^6$ times faster.}
    \label{fig:WEComp}
\end{figure}

Due to the inherent stochasticity of a surface CRN simulation, data generated these simulations is inherently noisy, even at a steady state. While the system may have settled into a state of relatively constant concentration, the inherent movement of species in the system can lead to small variations in overall species counts. As more species are added or as faster reactions are added to the regime relative the overall duration of the simulation, the number of concentrations calculated increases dramatically. This compounds the noisy data problem to be incredible dense, making sensitive calculations of values such as $K_{eq}$ and $Q$ inherently messy. In order to properly calculate systemic descriptors such as $K_{eq}$ and $Q$, chemical data must be smoothed in order to eliminate noise propagation in results. One method for addressing this is through the Whittaker-Eilers smoother, a smoother based on penalized least squares. Extremely fast compared to classic data smoothing techniques like the Savitzky-Golay filter and moving averages, the Whittaker-Eilers filter gives continuous smoothness control as well as automatic interpolation and fast leave-one-out cross-validation. Fig. \ref{fig:WEComp} compares the curve generation of the Whittaker-Eilers smoother on noisy $Q$ data, showing a marked reduction in data noise similar to that of a moving average calculation, albeit at a fraction of the time to calculate.
\newline
Given a set of noisy data $y$, there is a series $z$ that is believed to be the optimal smoothness of $y$. As $z$ increases in smoothness, the residual between $z$ and $y$ increases. This residual $\epsilon$ is calculated as:
\begin{align}
    \epsilon = \sum_i(y_i - z_i)^2
\end{align}
and the smoothness $s$ of the data is calculated as:
\begin{align}
    s = \sum_i (z_i - z_{i-1})^2 = \sum_i (\Delta z)^2
\end{align}
To balance the $\epsilon$ and $s$ is tuned by the user through smoothing parameter $\lambda$, with the relationship between this quantity represented as $q$:
\begin{align}
    q = \epsilon + \lambda s
\end{align}
Ultimately, the Whittaker-Eilers smoother finds the series $z$ that minimizes $q$. Combining the above expressions and defining $y$ and $z$ as vectors $\mathbf{y}$ and $\mathbf{z}$ as well as a differential matrix $D$, the expression for $q$ evolves to:
\begin{align}
    q = |\mathbf{y} - \mathbf{z}|^2 + \lambda |\mathbf{D}\mathbf{z}|^2
\end{align}
Minimizing $q$ via setting the gradient of $q$ to 0, we arrive at the following expression:
\begin{align}\label{Eq:WEexpression}
    (\mathbf{I} + \lambda \mathbf{D}^T\mathbf{D})\mathbf{z} = \mathbf{y}
\end{align}
with $\mathbf{I}$ defined as the identity matrix. Eq. \ref{Eq:WEexpression} is of the form $\mathbf{A}\mathbf{z} = \mathbf{y}$ and can thus be solved via matrix decomposition to find $\mathbf{z}$.
\newline

\subsection{Minkowski Functionals}

Minkowski functionals are geometric and topological descriptors derived from integral geometry used to describe spatial patterns\cite{Boelens2021-jz}. For a system of dimension $D$, $D+1$ functionals are required to describe it. In the case of a 2D system with a surface $\Omega$ and a smooth boundary $\delta\Omega$, the required functionals are defined as:
\begin{align}
    M_0(\Omega) &= \int_{\delta\Omega}dA \label{Eq:M0} \\
    M_1(\Omega) &= \frac{1}{2}\int_{\delta\Omega}dL \label{Eq:M1} \\
    M_2(\Omega) &= \frac{1}{2}\int_{\delta\Omega}k(\Omega)dL = \pi\chi \label{Eq:M2}
\end{align}
Where $dA$ is defined as a surface element, $dL$ is a line element, and $k(\Omega)$ is the signed curvature. For our purposes, $M_0$ defined as the surface area of the system, $M_1$ as the perimeter, and $M_2$ as the signed curvature, which is directly proportional to the Euler characteristic $\chi$ via the Gauss-Bonnett theorem \cite{Guevel2021-jf}. For any functional $\boldsymbol{M}(\Omega)$ that is additive, motion invariant and continuous, per Hadwiger's theorem \cite{Klain1995-nv}, this functional can be described as a linear combination of Minkowski functionals $M_n(\Omega)$ as follows:
\begin{align}\label{Eq:MinkGeneral}
    \boldsymbol{M}(\Omega) = \sum^d_{n = 0}c_nM_n(\Omega)
\end{align}



\subsection{$K_{eq}$ selection}
In order to benchmark the effect Minkowski functionals have on chemical reaction systems, a simple dissolution reaction as described in Eq. \ref{Eq:Dissrxn} was studied. For experimental reproducibility and observability purposes,the homogenized properties of the reaction on a given microstructure have to be evaluated at the reaction's steady-state. At this point,the reaction is classically characterized through a measure of the total extent of the reaction measured through the reaction quotient $Q_r$ and its value at steady-state called the equilibrium constant $K_{eq}$, as well as a measure of its duration measured through its equivalent reaction rate/constant. These two are going to be the properties that the present work will focus on as well, starting from defining a representative form of $K_{eq}$ for non-mixing systems.

In classical mixing systems, the forms of $K_{eq}$ and $Q_r$ for this dissolution reaction are assumed to be derived from the law of mass action, as follows:
\begin{align}
\label{eq:KeqR2}
    K_{eq} &= \frac{[R_{eq}]^{2}}{[Q_{eq}][A_{eq}]} \\
    Q_r &= \frac{[R]^{2}}{[Q][A]}
\end{align}
Equilibrium values used to calculate $K_{eq}$ take the mean of the last few values of the system at steady state, reducing the overall noise for calculations. However, the ultimate results of this assumption comes into question, specifically in the agreement of $K_{eq}$ results relative to the work proposed by Boelens et al\cite{Boelens2021-jz}. From this work, an agreement of the following form is expected based on classical (additive) concepts of thermodynamics:
\begin{align}
    \Delta G = -RT\ln K_{eq} = \alpha M_0 + \beta M_1 + \gamma M_2
\end{align}
Based on this relation, an additive, linear combination of Minkowski functionals in an exponential distribution is traditionally expected to describe the energetics of the system. 

However, non-mixing systems have been shown to deviate from the law of mass action for over 50 years\cite{Haase}. 
%
The burgeoning work of surface chemistry energetics has resurfaced these considerations, suggesting that the traditional law of mass action described in Eqns. \ref{eq:KeqR2} is not accurate in systems with multiple state phases\cite{Bauermann2022-va, Drobot2018-ra, Nakashima2019-bm}. Bauermann et al. specifically defines $K_{eq}$ as a relationship between stroichiometric coefficients 
, activity coefficients 
, and reference chemical potentials suggesting slower versions of $K_{eq}$ for non-mixing interface reactions based on these metrics.
Unfortunately, in this synthetic system these considerations are not exactly applicable since energetic terms like chemical potential and activity coefficients are assigned a priori in the form of transition rule rates and diffusion rates, respectively. As a result we can only homogenize numerically and to this extent three $K_{eq}$ formulations will be examined-- those from Eqns. \ref{eq:KeqR2} and two slower versions defined as
\begin{align}\label{eq:KeqR32}
    K_{eq}^{R^{n}} &= \frac{[R_{eq}]^{n}}{[Q_{eq}][A_{eq}]}, \: n=1; \: 3/2; \: 2
\end{align}
The contrasting results from these varying $K^{R^{n}}_{eq}$ calculations will inform an ultimate selection for the $K_{eq}$ criteria that Minkowski functional analysis will be based on. Further to the extent of the reaction, its equivalent rate will be represented through a characteristic time of the reaction, $\Delta \tau$, that will in turn be investigated as a function of Minkowski functionals of an assumed form:
\begin{align}
     \Delta \tau = f(c_nM_n)
\end{align}

\section{Numerical Results on synthetic microstructures}
\subsection{Microstructure Selection}
One of the challenges with testing the localized effects of individual Minkowski functionals on microstructural properties is the inherent difficulty in isolating Minkowski functionals from each other in microstructural generation. Indeed, while these functionals are by definition linearly independent, it is quite difficult creating a schema that only varies one functional while fixing the others. In order to isolate individual Minkowski functionals, three microstructural designs were created.
\newline

\begin{figure}[ht!]
    \centering
    \includegraphics[scale=0.5]{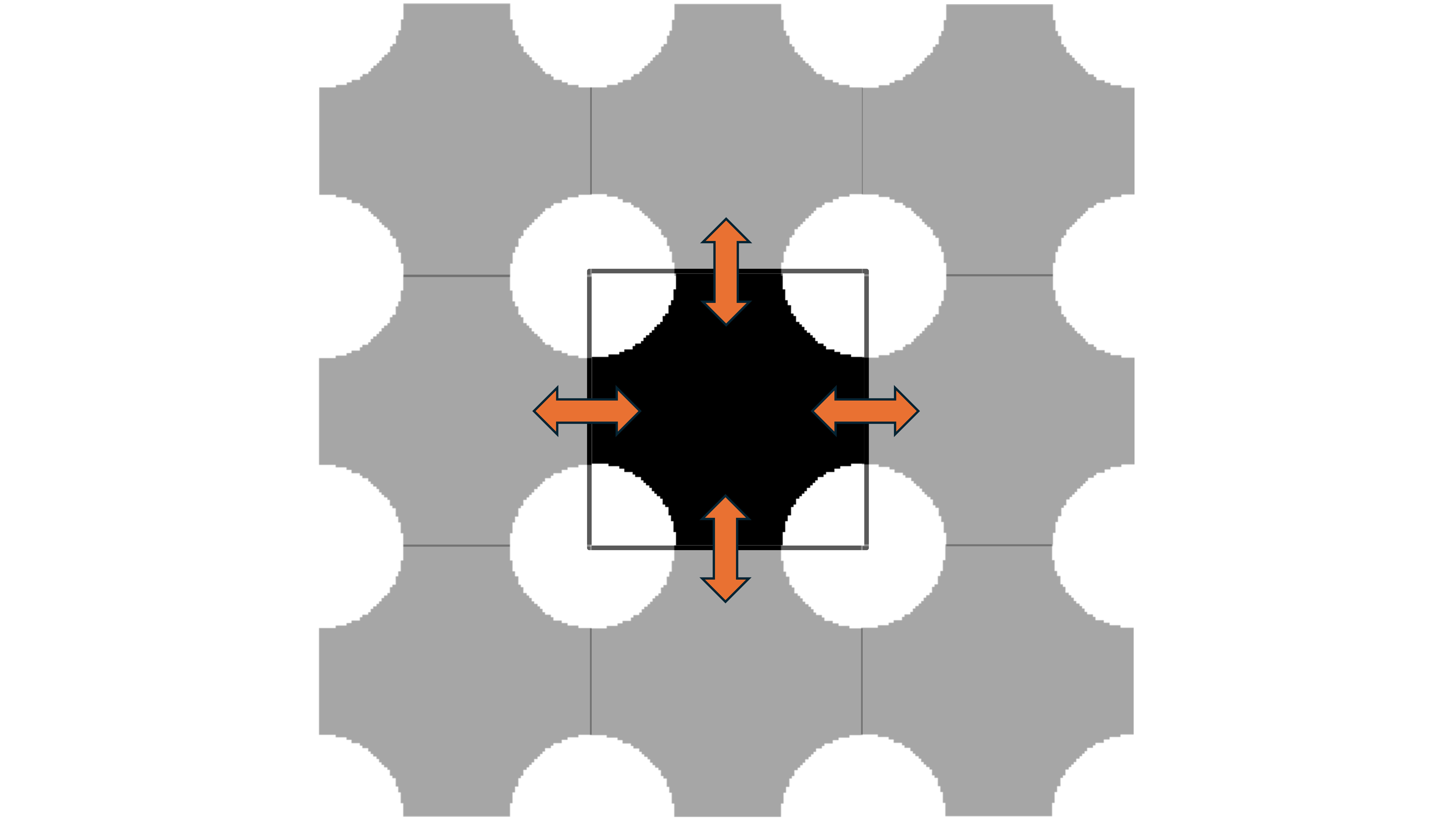}
    \caption{Example of a unit cell microstructure. The radius of the circles at the corners are varied per individual unit cell designs. The boundary of the unit cells are periodic, allowing for chemical reactions to occur from one edge to another.}
    \label{fig:UnitCellDesign}
\end{figure}

\begin{figure}[ht!]
    \centering
    \includegraphics[scale=0.5]{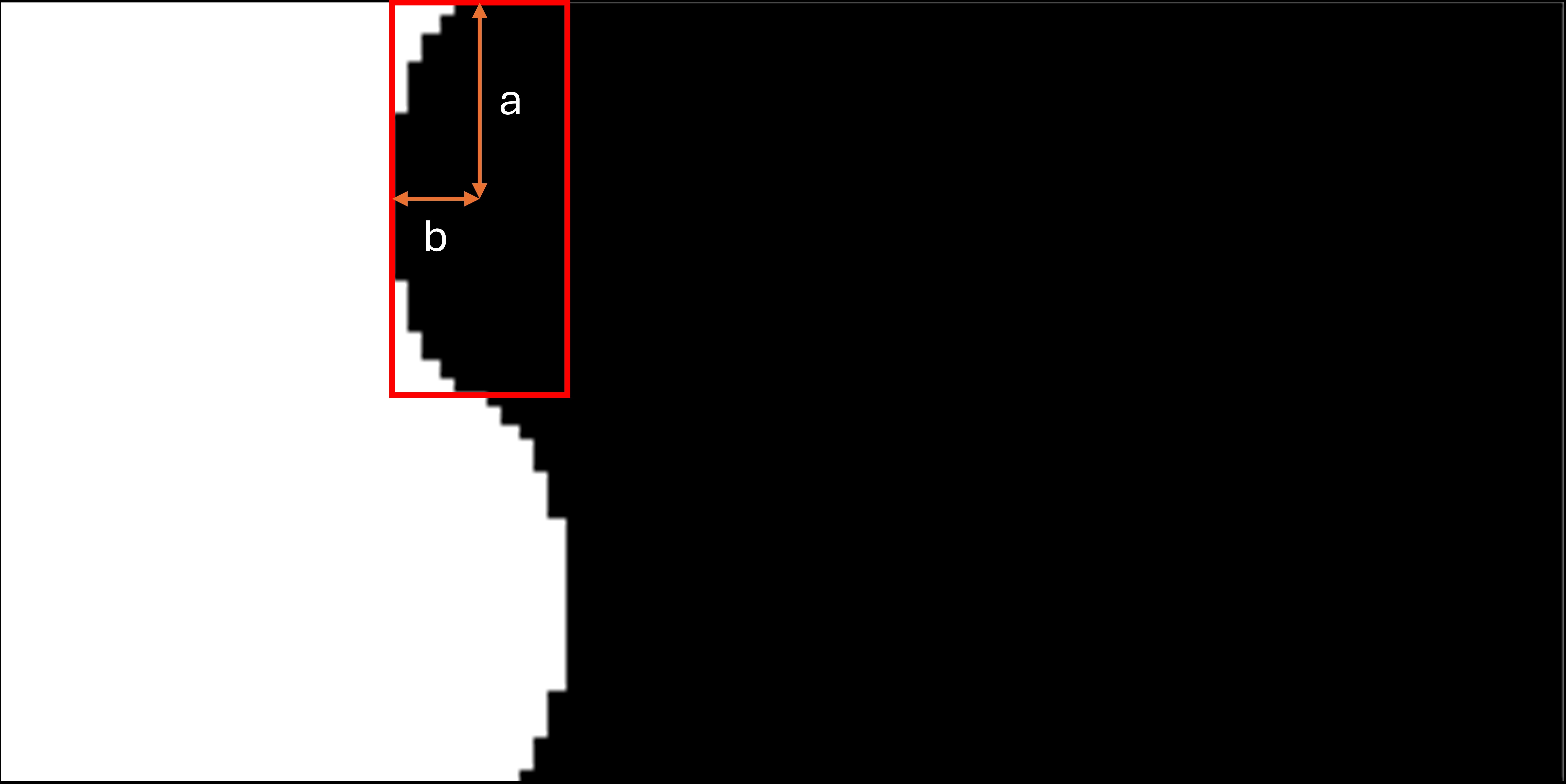}
    \caption{Example of a perimeter test microstructure. $a$ and $b$ control the wave properties along the perimeter, varying $M_1$ while maintaining a constant $M_0$. One periodic wave is highlighted in the red bounding box, with the wave number of the cell defined as $\nu$.}
    \label{fig:PerimeterDesign}
\end{figure}

Fig. \ref{fig:UnitCellDesign} demonstrates the first microstructural design, a periodic unit cell. Each unit cell is designed as an $N$x$N$ square with four circles of equal radius $r$ at each corner. White pixels represent solid species while black pixels represent voids for fluid species to diffuse. Each edge of the unit cell is a periodic boundary, allowing for chemical reactions to occur from one end of the cell to the other. To generate unit cells of differing Minkowski functionals, the unit cell bounding box is fixed at side length $N$ as $r$ is varied. While Eqs. \ref{Eq:M0}-\ref{Eq:M2} hold as the basis for calculating Minkowski functional values, $M_0$ and $M_1$ are nondimensionalized by the reference length $N$ of the bounding box. Thus, Minkowski functionals are calculated as follows:
\begin{align}
    M_0 &= 1 - \frac{\pi r^2}{N^2} \label{Eq:M0porosity} \\
    M_1 &= \frac{2\pi r}{N} \label{Eq:M1perimeter} \\
    M_2 &= V - E + F \label{Eq:M2EC}
\end{align}
With $V$, $E$, and $F$ of Eq. \ref{Eq:M2EC} representing the vertices, edges, and faces of the microstructure, respectively. Note that Eq. \ref{Eq:M0porosity} is calculated as the porosity of the system (fraction of void to total box size). For unit cell tests, $M_2$ is held constant ($\chi = 1$ for a circle split into four slices) while $M_0$ and $M_1$ vary with $r$.
\newline
In order to separate the effects of $M_0$ and $M_1$, a second test was designed to hold $M_0$ and $M_2$ constant while only varying $M_1$. In Fig. \ref{fig:PerimeterDesign}, an example of a perimeter test microstructure is shown. In order to hold a constant $M_0$ and $M_2$, a boundary on a solid region of the microstructure has a periodic wave applied to it, varying perimeter while keeping the same area ratio from the solid to fluid area. The number of waves on the perimeter is denoted by the wave number $\nu$. The perimeter and area of the wave boundary is calculated in a similar manner to that of an ellipse, thus $a$ and $b$ represent shape measures for calculating wave area and perimeter. Because of the periodic nature of the wave boundary, $M_0$ and $M_2$ remain constant while $a$, $b$, and $\nu$ are varied (assuming $\nu$ remains an even number). Based on the Ramanujan approximation for the perimeter of an ellipse, $M_1$ is calculated as:
\begin{align}
\begin{split}
    M_1 &= \frac{\nu\pi(a + b)}{2}\Bigg(1+\frac{3h}{10 + \sqrt{4-3h}}\Bigg) \\
    h &= \frac{(a - b)^2}{(a + b)^2}
\end{split}
\end{align}

\begin{figure}[ht!]
    \centering
    \includegraphics[scale=0.5]{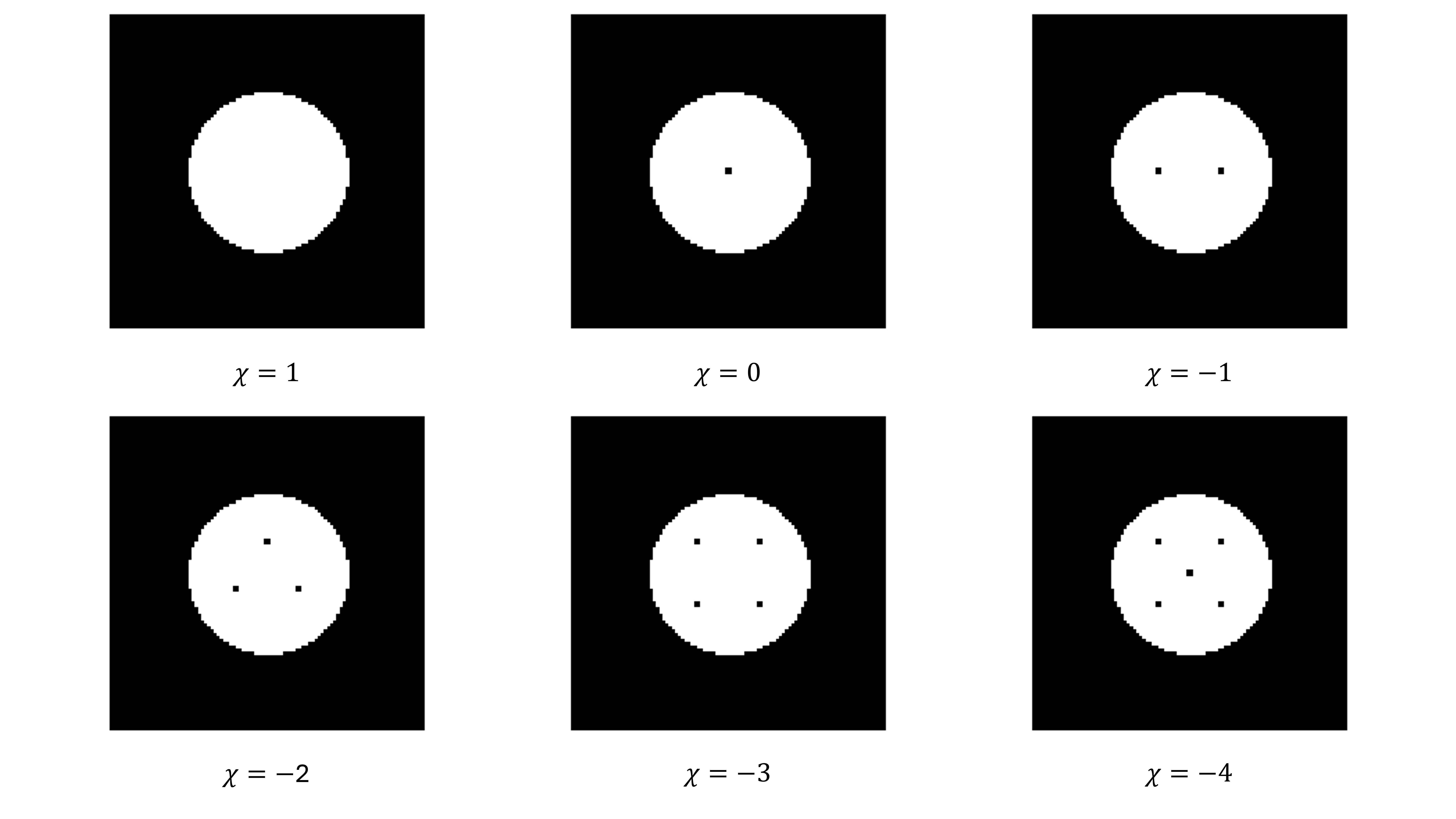}
    \caption{Euler characteristic test. All microstructures are circles of constant radius, with pixel-sized holes added. Each hole lowers $\chi$ by 1, at a negligible change in porosity and perimeter.}
    \label{fig:ECDesign}
\end{figure}

The final microstructural design aims to maintain a constant $M_0$ and $M_1$ while varying $M_2$. Fig. \ref{fig:ECDesign} represents how this test was performed, with a circle of solid material immersed in a bounding cell of fluid. As pixel-sized holes are added to the circle, $\chi$, and therefore $M_2$, decreases. Due to the small size of these holes, $M_0$ and $M_1$ change negligibly through the test.

\subsection{Resolution Convergence}
\begin{figure}
    \centering
    \includegraphics{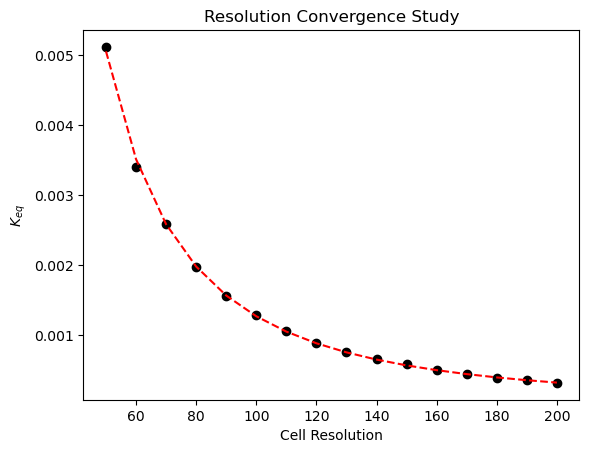}
    \caption{Resolution convergence study, varying unit box size.}
    \label{fig:resolutionconvergence}
\end{figure}

In order to assess the performance of surface CRNs as a modeling tool for chemical behavior, a resolution convergence study was performed to check if $K_{eq}$ values scaled directly with simulation resolution. If resolution showed to have a considerable effect on model performance, special consideration would have to be made when considering the scaling dimensions of a simulation.
\newline
To perform a resolution convergence study, repeated simulations were performed on the periodic unit cell, varying cell side length $N$ while keeping the ratio of side length to circle radius $r$ at a consistent 4:1 $N:r$ ratio. $K_{eq}^R$ was selected as the convergence criteria, allowing for steady state solutions to be the sole source of comparison, as seen in Fig. \ref{fig:resolutionconvergence}. By solely comparing steady-state values, the effects of cell resolution can be investigated on their own. Due to increased resolution, dynamic effects in the unit cell would need to be scaled via the transition rule rate laws, as the increased resolution would effectively increase the "distance" each set of molecules would need to travel due to the fixed grid nature of surface CRN simulations.

As seen in Fig. \ref{fig:resolutionconvergence}, $K_{eq}$ values show a clear exponential decrease with increasing resolution, converging at a consistent solution at about $N = 200$.

\subsection{Rate Effects}\label{sec:rateeffects}
\begin{figure}
    \centering
    \includegraphics{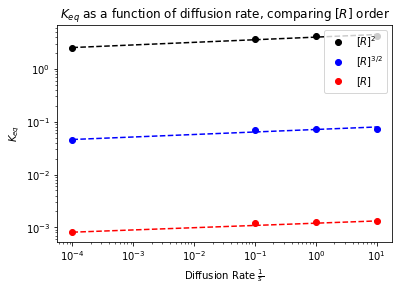}
    \caption{The effect of diffusion rate on the $K_{eq}$ of the system for each $K_{eq}$ formulation.}
    \label{fig:DiffusionStudy}
\end{figure}

According to the work of Boelens et al.\cite{Boelens2021-jz}, the primary discrepancy in $K_{eq}$ values found in interfacial systems compared to well-mixed systems manifests from differing reaction rates, both within the separate phases but also in the transition from one phase to another. In surface CRN simulations, these discrepancies can manifest in the a priori transition rule rates, as well as the assigned diffusion rate for "motile" species in the simulation space. Fig. \ref{fig:DiffusionStudy} demonstrates how an increasing diffusion rate increases $K_{eq}$ consistently across varying methods for $K_{eq}$ calculation. These increases all matched closely to a power law, with consistent power scaling across all three calculation schemes. The primary difference in each curve stems from the order of magnitude of $[R]$ at a consistent linear scaling.

\begin{figure}
    \centering
    \includegraphics{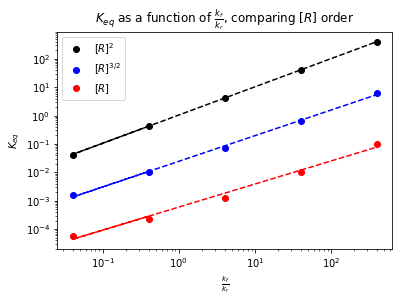}
    \caption{The effect on reaction rate ratio for the forward and reverse reaction on $K_{eq}$ for each $K_{eq}$ formulation.}
    \label{fig:reactionrate}
\end{figure}

A similar class of study was performed comparing the rate of reaction in transition rules. As detailed in Eq. \ref{Eq:Dissrxn}, the benchmark reaction is a reversible reaction that in its outset favors a forward reaction. For this study, the ratio of the forward reaction $k_f$ to the reverse reaction $k_r$ was varied, as in Fig. \ref{fig:reactionrate}. Similar to the behavior exhibited in Fig. \ref{fig:DiffusionStudy}, $K_{eq}$ calculations varied consistently across the same order power law, modulating by constant orders of magnitude per the $K_{eq}$ formulation.
\newline
Both rate effect studies shared consistent results in terms of the scalability of $K_{eq}$ calculations across various rate schemes and diffusion rules. The influence of these varying rates points to the validity of Boelens' work, as the kinetics of the varying phases of the reaction, both chemically and physically, have a direct influence on the overall steady state behavior of the system.

\subsection{Interface Diffusion Phenomenon as Internal Branching}

While the kinetics of the system have direct, tangible effects on the overall behavior of the $K_{eq}$ calculation, another important area of consideration is the idiosyncrasies of the simulation medium used in this study. While surface CRNs possess inherent advantages compared to other discrete stochastic simulators in their inherent spatiality and simple solving scheme, secondary behavior may arise depending on the nature of the reaction rules given to the system. In the case of this reaction, a slow but noticeable phenomenon of diffusion was observed to occur, even in systems where no diffusion amongst fluids was prescribed.

\begin{figure}
    \centering
    \includegraphics[scale=0.45]{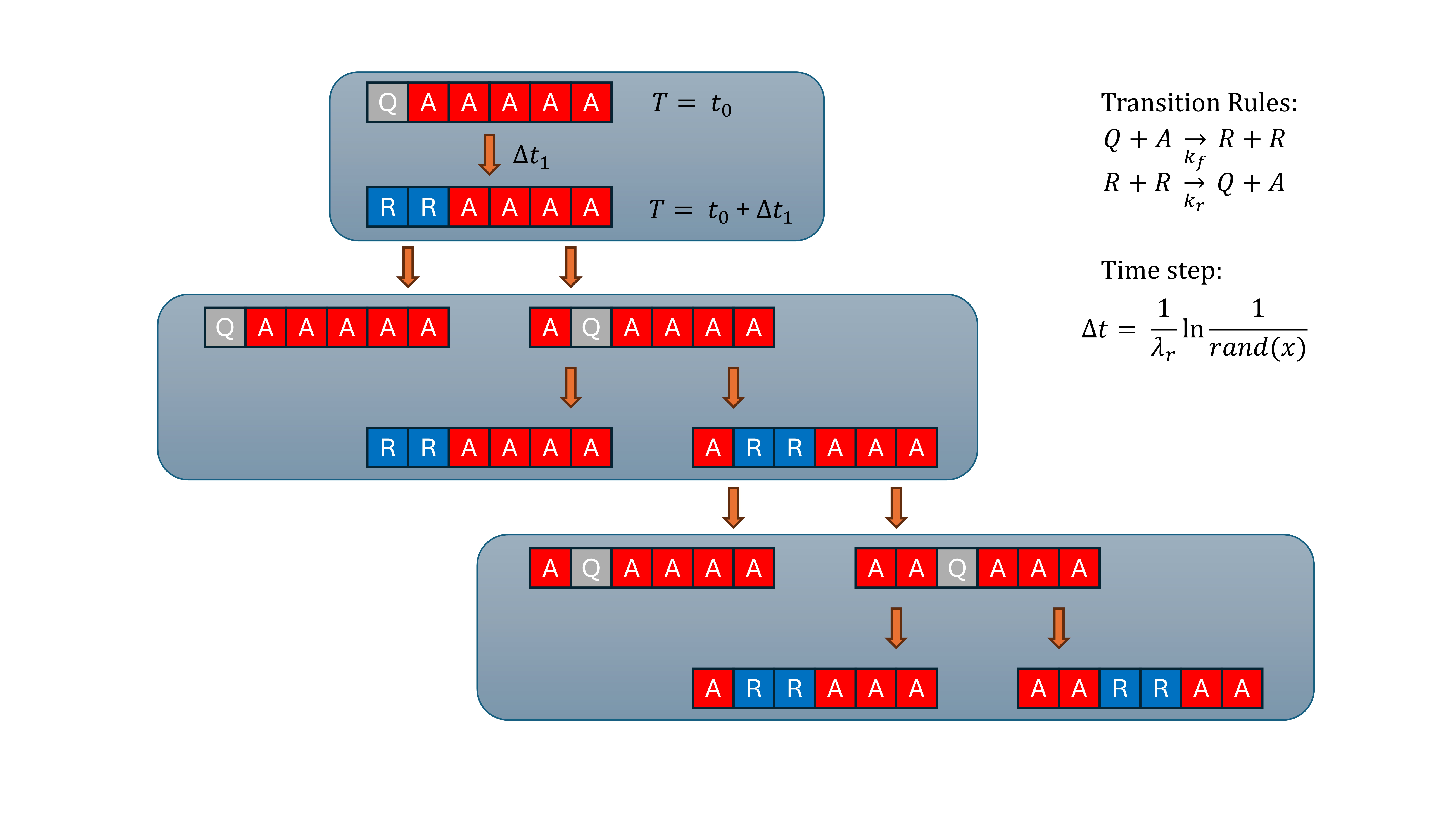}
    \caption{Diagramatic description of branching diffusion.}
    \label{fig:branchingdiffusion}
\end{figure}

\begin{figure}
    \centering
    \includegraphics[scale=0.45]{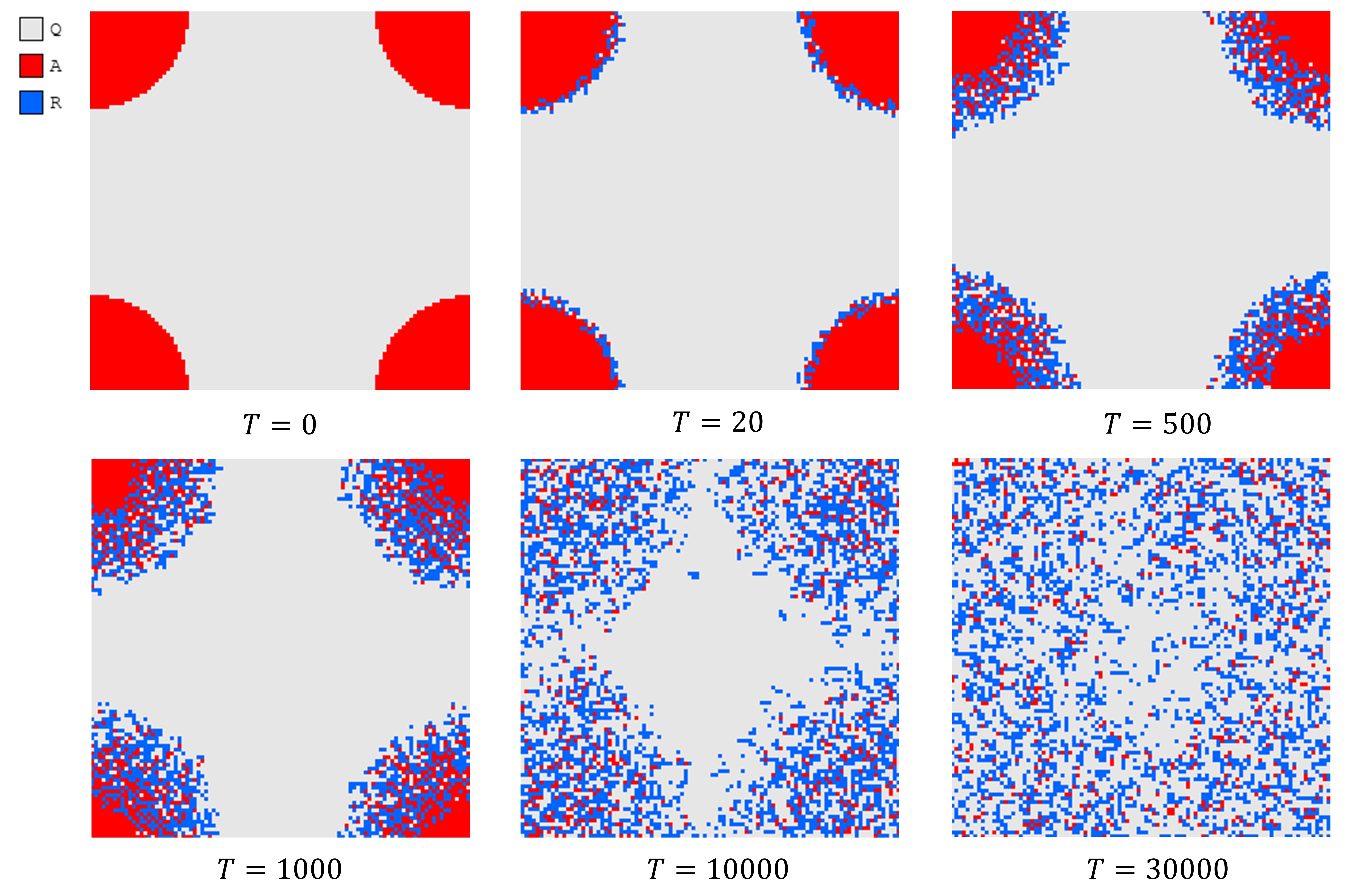}
    \caption{Time evolution of interface diffusion across a periodic unit cell.}
    \label{fig:interfacediffusion}
\end{figure}

Typically, in a system where diffusion is disabled (in our case, the rate of the diffusion transition rule is set to 0), chemical reactions occur almost instantaneously at the solid-fluid interface and then stop, creating a layer of product at the boundary. This is because without some form of transport, reacting species may only form a layer at the surface boundary before the subsequent product shields further reactions from occurring, ultimately terminating the surface CRN simulation early due to the reaction queue collapsing. However, in the case of the class of reaction discussed in this work, the fact that the product of the reversible reaction is two of the same species creates a unique scenario where a slow diffusion manifold is allowed to propagate in the system. Diagrammed in Fig. \ref{fig:branchingdiffusion}, this slow self-propagating diffusion manifold--or internal branching diffusion--is tied directly to the probabilistic nature of asynchronous cellular automata.
\newline
After the initiation of a chemical reaction on a surface that generates two identical product species with the contact of a reactive fluid $Q$ in the presence of a reactive solid $A$, the surface CRN faces a conundrum for its next step--how to resolve the two identical species with the potential for the reverse reaction. As the surface CRN scans each newly generated node species for potential chemical reactions, it finds that both $R$ product species are eligible for a subsequent chemical reaction to occur. Thus, both reaction sites draw a random $\Delta t$ that dictates which of the two sites initiates a reaction first. Depending on which site draws a faster reaction--of which both sites have an equal probability of this occurring--the reverse reaction may assign either site to revert to either $Q$ or $A$. If this reversibility goes back to the direction of the initial propagation, the reaction oscillates at the boundary between products and reactants. However, if the order of the $Q$ and $A$ reactive sites flip, the dynamics of the reactions change as now there are sites inside of the solid past the initial boundary that are in contact with reactive nodes. At each step of this flip occuring, new internal potential reaction sites are exposed, propagating the initial reaction through the solid phase. This effect is bidirectional, as these flips may occur in the other direction to move species at the original solid-fluid boundary outward, essentially mirroring a slow diffusion process. With these dynamics incorporated in the system, even with no diffusion prescribed in the transition rules, the ultimate fate of the system at steady state eventually sees the entire solid state dissolve into product, with product dispersed evenly throughout the reacting cell as seen in Fig. \ref{fig:interfacediffusion}.
\newline
While this phenomenon occurs at an incredibly slow rate, with convergence to steady state occurring orders of magnitude further than systems with even the slowest diffusion constant, this slow manifold directly influences the rate and availability of the reactions in this chemical system. This small-scale diffusion phenomenon absolutely has the potential to influence the ultimate $K_{eq}$ behavior of the system, further lending discrepancies in $K_{eq}$ in well mixed systems versus those in interfacial systems. 

\section{Homogenization of Microstructure Geometry}
\subsection{Unit Cell}

\begin{figure}
    \centering
    \includegraphics[scale=0.45]{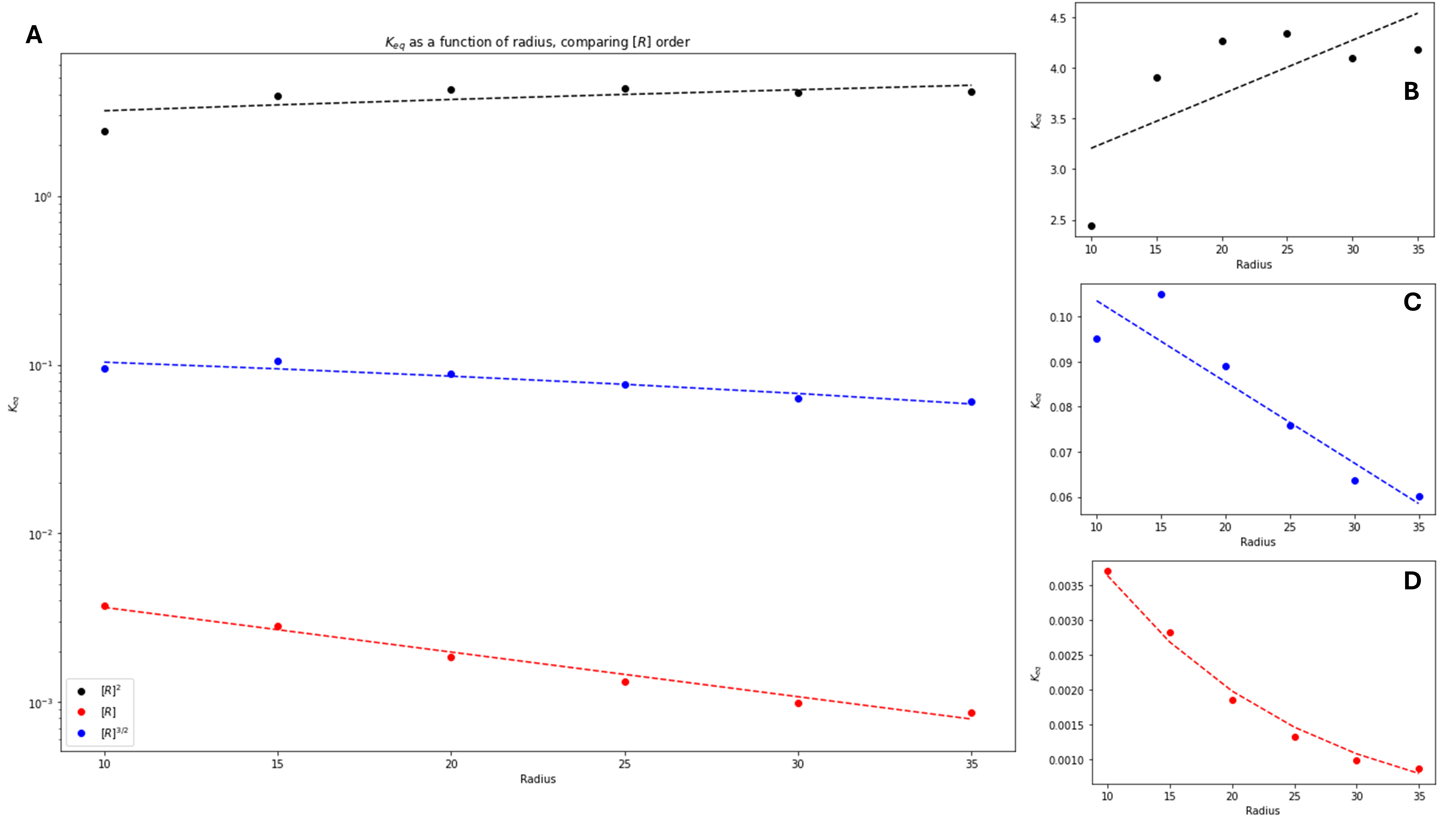}
    \caption{Comparing the effect of radius on various $K_{eq}$ calculation schemes. A) compares across all three schemes of $K_{eq}^R$, $K_{eq}^{R^{3/2}}$, $K_{eq}^{R^2}$, B) plots $K_{eq}^{R^2}$ as a function of radius, C) plots $K_{eq}^{R^{3/2}}$ as a function of radius, and C) plots $K_{eq}^R$ as a function of radius.}
    \label{fig:KeqCompRadius}
\end{figure}

As discussed above, when investigating the effects Minkowski functionals have on the chemical performance of the system, a clear definition of $K_{eq}$ must be defined. In Fig. \ref{fig:KeqCompRadius}, it is clear that depending on the selected scheme of calculating $K_{eq}$, the reference scaling and relational behavior with regards to radius changes dramatically. Fig. \ref{fig:KeqCompRadius}B shows a weak linear, bordering on trivial, relationship between radius and $K_{eq}^R$. Fig. \ref{fig:KeqCompRadius}C on the other hand shows a strong linear relationship between radius and $K_{eq}^{R^{3/2}}$. Finally, Fig. \ref{fig:KeqCompRadius}D shows a strong exponential relationship between radius and $K_{eq}^R$. The differentiation in these schemes only appears in the calculation of $K_{eq}$ itself, and not in other kinetics related factors such as $\Delta \tau$.

\begin{figure}[ht!]
    \centering
    \includegraphics[scale=0.5]{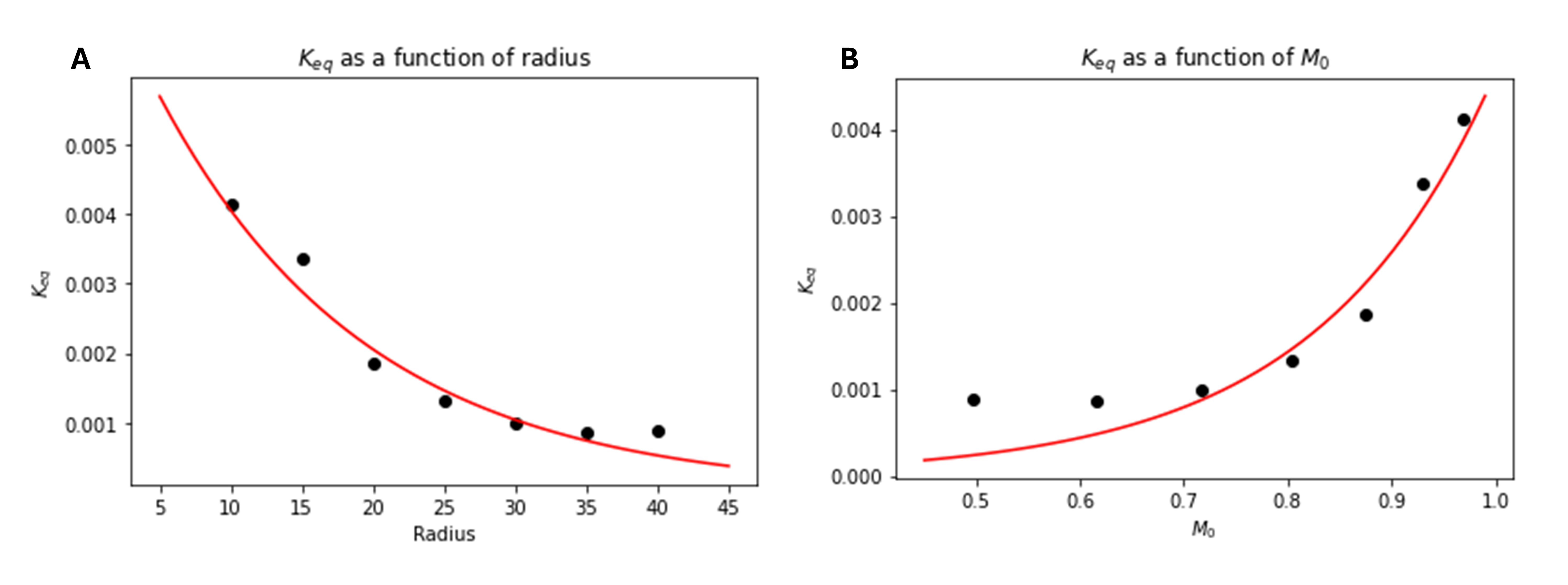}
    \caption{$K_{eq}^R$ values for unit cell reactions of a benchmark dissolution reaction. A) plots the evolution of $K_{eq}^R$ as a function of unit cell radius. B) plots the evolution of $K_{eq}^R$ as a function of $M_0$.}
    \label{fig:UCDissolutionQ}
\end{figure}

\begin{figure}[ht!]
    \centering
    \includegraphics[scale=0.45]{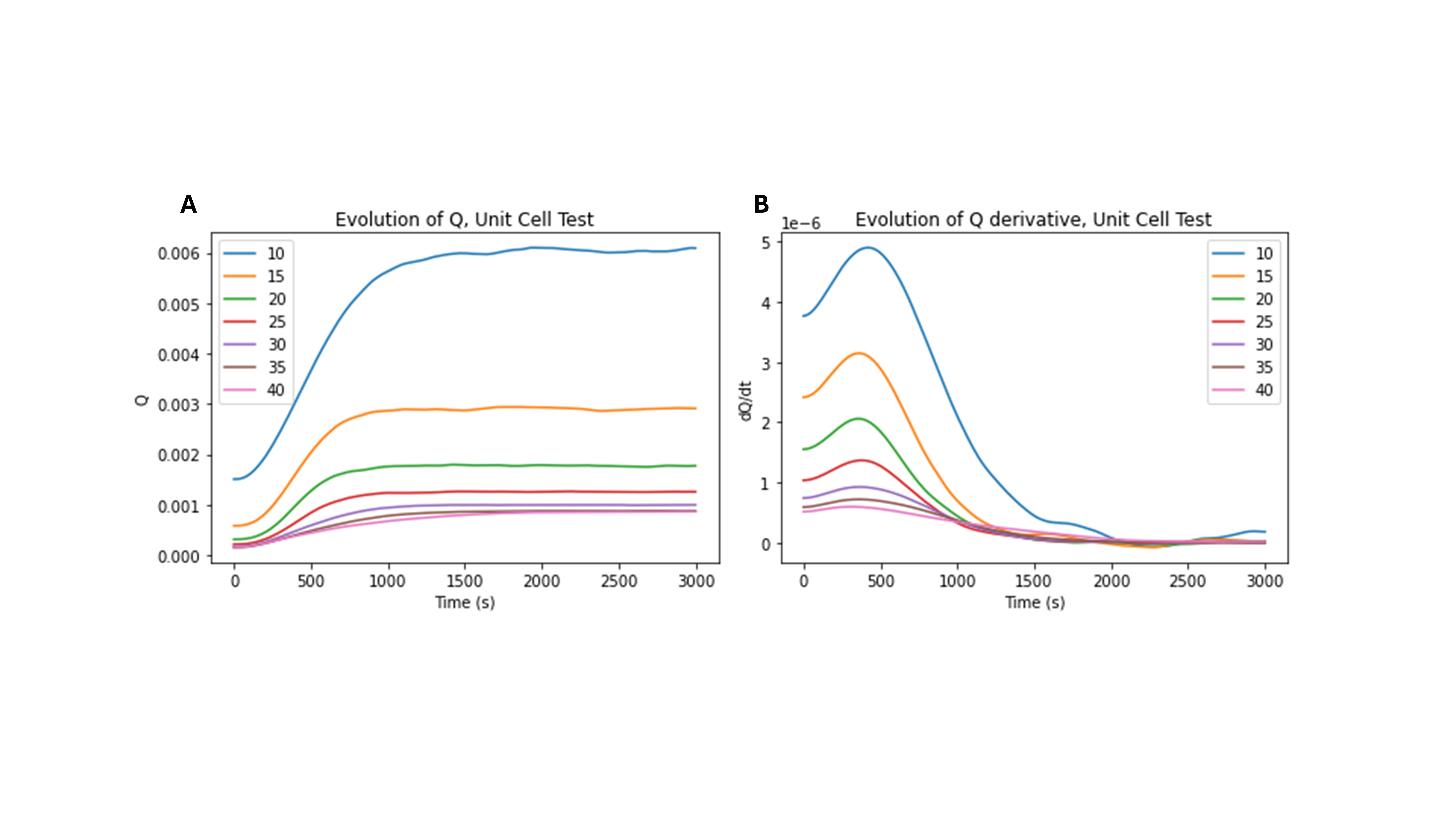}
    \caption{$Q^R$ values for unit cell reactions of a benchmark dissolution reaction. A) plots the evolution of $Q^R$ as a function of unit cell radius over time for all test systems. B) plots the evolution of the first derivative of $Q^R$, $\frac{dQ^R}{dt}$ as a function of unit cell radius.}
    \label{fig:UCDissolution}
\end{figure}

In examining the unit cell behavior of the benchmark dissolution reaction, a range of radii from 10 to 40 were tested. In terms of $M_0$, this resulted in a porosity range from 0.5 to 0.95. As shown in Fig. \ref{fig:UCDissolution}, an exponential relationship was found between the terms and $K_eq^R$, with a negative exponential found relating to radius, and a positive exponential found with respect to $M_0$. The greatest error in the plot can be found at the largest radius (and thus, lowest porosity) of the unit cell. This is likely due to the chemical bearing capacity of the unit cell itself--without a transport means for chemical species to exit the unit cell system, lower porosity unit cells likely experience greater chemical exclusion effects due to the spatial nature of surface CRNs.

When plotting the entire $Q^R$ profile, as seen in Fig. \ref{fig:UCDissolutionQ}, the exponential relationship between the radius of the unit cell and the steady state of the system is made clear. There is also a clear relationship between the maximum $\frac{dQ^R}{dt}$ and the overall radius of the circles in the unit cell. Notably, while the maximum $\frac{dQ^R}{dt}$ varies significantly with radius and porosity, there is less clear of a trend in the time to reach the maximum $\frac{dQ^R}{dt}$, or $\Delta \tau$. Indeed, outside of edge cases (highest radius and lowest radius), $\Delta \tau$ appears to be largely unchanged.

\begin{figure}[ht!]
    \centering
    \includegraphics[scale=0.65]{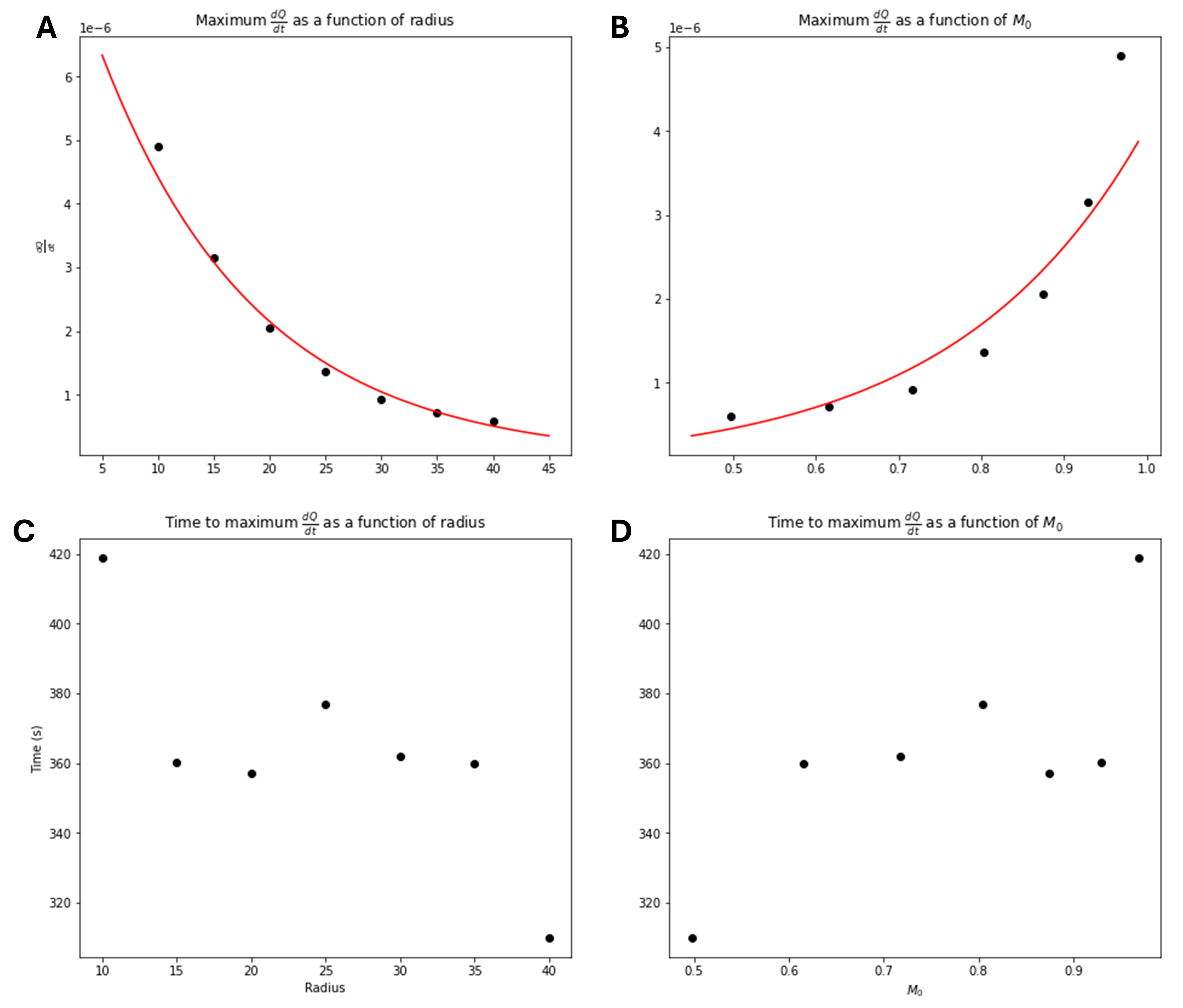}
    \caption{$\frac{dQ^R}{dt}$ values for unit cell reactions of a benchmark dissolution reaction. A) plots the maximum $\frac{dQ^R}{dt}$ as a function of unit cell radius. B) plots $\frac{dQ}{dt}$ as a function of unit cell $M_0$. C) plots the time to maximum rate $\Delta \tau$ as a function of radius. D) plots $\Delta \tau$ as a function of $M_0$.}
    \label{fig:UCDissolutiondQdt}
\end{figure}

\subsection{Perimeter}

Perimeter was varied as was described in the section above, with both wave parameters $a$ and $b$ varied to generate different perimeter values for the testing sample, as seen in Fig. \ref{fig:PerimeterDesign}. Ultimately these results were combined to draw overall conclusions surrounding the effect of $M_1$ on microstructural chemical performance.

\begin{figure}
    \centering
    \includegraphics[scale=0.55]{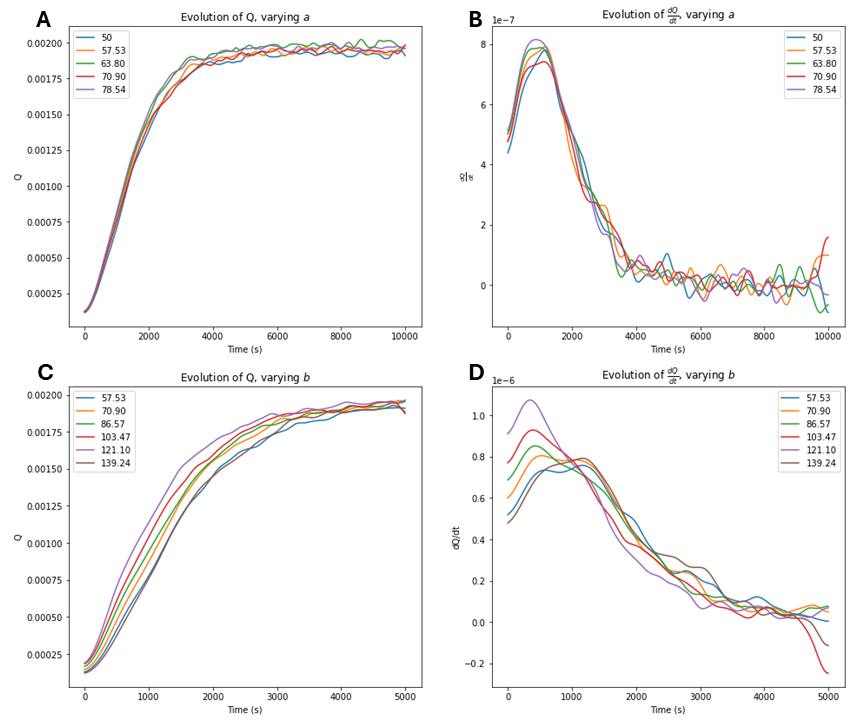}
    \caption{Plots of $Q^R$ evolution through varying perimeter tests. A) tracks the overall evolution of $Q^R$ at different perimeters, controlled by varying $a$. B) tracks $\frac{dQ}{dt}$ at different perimeters, also through varying $a$. C) tracks the overall evolution of $Q^R$ at different perimeters, controlled by varying $b$. D) tracks $\frac{dQ}{dt}$ at different perimeters, also through varying $b$.}
    \label{fig:PerimeterQPlots}
\end{figure}

\begin{figure}
    \centering
    \includegraphics[scale=0.45]{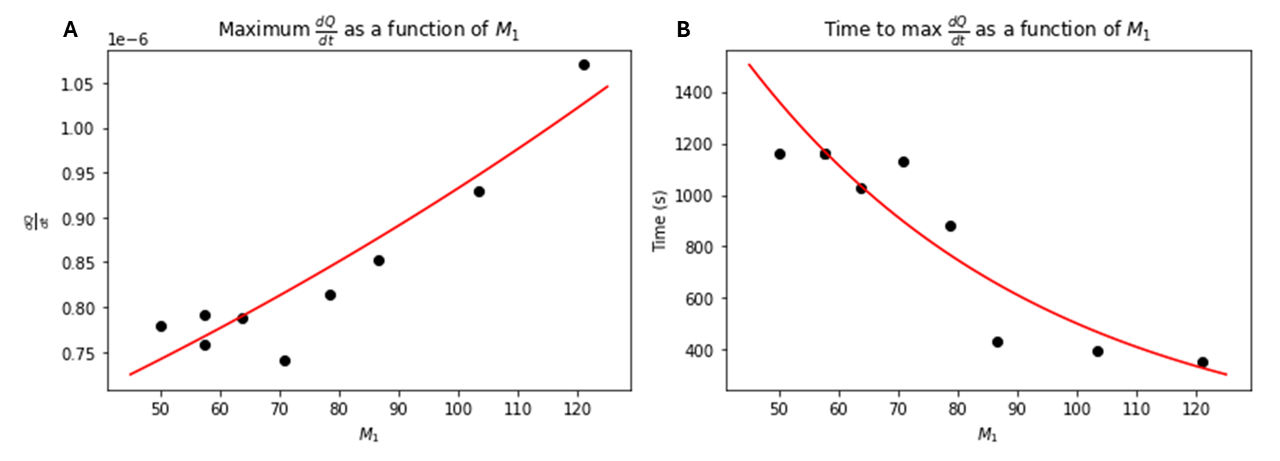}
    \caption{Effect of $M_1$ on $\frac{dQ}{dt}$. A) demonstrates the exponential relationship between $M_1$ and the maximum $\frac{dQ}{dt}$ of the system. B) demonstrates the exponential relationship between $M_1$ and the time to maximum $\frac{dQ}{dt}$.}
    \label{fig:PerimeterRates}
\end{figure}

From Fig. \ref{fig:PerimeterQPlots}, in all cases of perimeter, $K_{eq}$ is unchanged outside of minor fluctuations expected of the stochastic nature of Surface CRN experiments. However, in the $\frac{dQ}{dt}$ plots in Fig. \ref{fig:PerimeterQPlots}B and D, clear hierarchy is seen through the relationship of perimeter to $\frac{dQ}{dt}$ behavior. This is further examined in Fig. \ref{fig:PerimeterRates}, where the relationship between $M_1$ and the maximum $\frac{dQ}{dt}$ and time to maximum $\frac{dQ}{dt}$ ($\Delta \tau$) is examined. In both cases, an exponential relationship is derived, although the relationship between $M_1$ and maximum $\frac{dQ}{dt}$ is positive exponential while the relationship between $M_1$ and $\Delta \tau$ is negative exponential.

\subsection{Euler Characteristic}

\begin{figure}[ht!]
    \centering
    \includegraphics[scale=0.5]{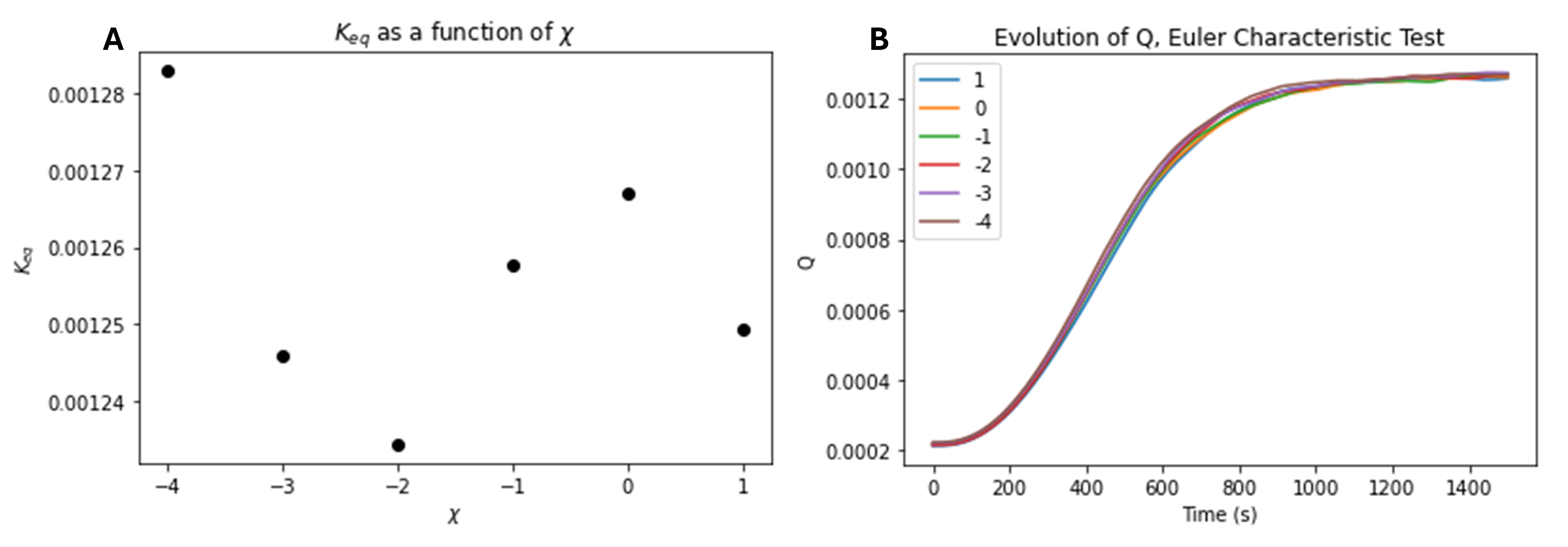}
    \caption{$K_{eq}^R$ and $Q^R$ values for Euler characteristic $\chi$ test of a benchmark dissolution reaction. A) Shows no relationship between $K_{eq}^R$ and $\chi$, while B) corroborates for the overall profile of $Q^R$.}
    \label{fig:ECKeq}
\end{figure}

As in the preceding sections, the benchmark chemical reaction from Eq. \ref{Eq:Dissrxn} was applied to the Euler characteristic $\chi$ testing scheme described in the methods section. Fig. \ref{fig:ECKeq}A demonstrates that $K_{eq}^R$ of the reaction system seems largely unaffected by the variations of the Euler characteristic. This is further corroborated in \ref{fig:ECKeq}B, where the overall $Q$ profile of each test varies minimally as $\chi$ changes.

\begin{figure}
    \centering
    \includegraphics[scale=0.5]{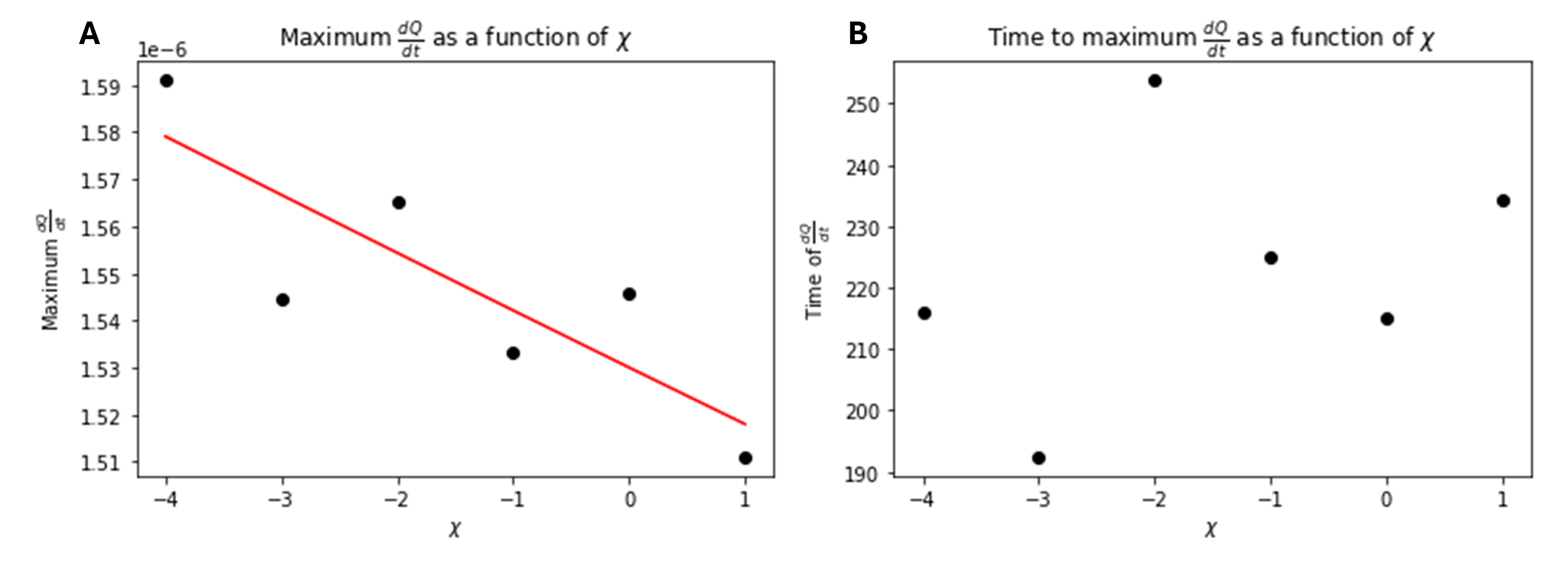}
    \caption{Rate effects of the $Q^R$ response from $\chi$. A) demonstrates the relationship between $\chi$ and $\frac{dQ}{dt}$. B) demonstrates the relationship between $\chi$ and $\Delta \tau$.}
    \label{fig:enter-label}
\end{figure}

While there is no clear relationship between $K_{eq}^R$ and $\chi$, a negative exponential relationship was observed between $\chi$ and the maximum $\frac{dQ}{dt}$ of the system. There was no visible relationship between $\chi$ and $\Delta \tau$.

\subsection{Dependency of reaction to morphometers}

From the data extracted from the various unit tests, the following table of relationships between morphometers and chemical reaction properties were extracted:

\begin{table}[]
\begin{tabular}{|l|l|l|l|}
\hline
                      & $K_{eq}^R$ & $\frac{dQ}{dt}_{max}$ & $\Delta \tau$ \\ \hline
Relevant Morphometers & $M_0$    & $M_0$, $M_1$, $M_2$   & $M_0$, $M_1$  \\ \hline
\end{tabular}
\end{table}

In all cases, exponential relationships were found. Ultimately, the only morphometer with a direct, tangible impact on the $K_{eq}$ was $M_0$. This is likely due to adjusting $M_0$ modulating the ratio of reactants in the system (i.e., a greater $M_0$ would decrease the amount of reactive solid $A$ and increase the amount of reactive fluid $Q$). However, while $M_1$ and $M_2$ had a minimal effect on the equilibrium behavior of the system, both functionals affected the dynamics of the system--$\frac{dQ}{dt}_{max}$ and $\Delta \tau$. These effects are likely due to $M_1$ and $M_2$ dictating the number of available reaction sites available in a system--perimeter determines the number of potential interfacial nodes while $\chi$ is a measure of the topological connectivity of the solid portion of the system. In both cases for these dynamic measures, $M_0$ would be relevant simply for adding more potentially reactive sites into the system.

\subsection{Linkage to Gibbs free energy}
With the simulation scaling in mind, the following mathematical relationships can be derived. For the equilibrium definition of Gibbs' free energy $\Delta G^o$, the following relationship is defined:
\begin{align}
    \Delta G^o = -RT\ln K_{eq}
\end{align}
When examining the relationships between Minkowski functionals and $K_{eq}$, we can define $K_{eq}$ as a function of $M_0$ in the following form:
\begin{align}
    K_{eq} = K_0 e^{a M_0}
\end{align}
for some constants $K_0$ and $\alpha$. In a similar vein, $\Delta \tau$ can be seen as a function of $M_0$ and $M_1$. This is of the form:
\begin{align}
\begin{split}
    \Delta \tau &= \Delta z_0(M_1) e^{b M_0} \\
    \Delta z_0 &= \Delta z_1 e^{c M_1} \\
    \Delta \tau &= \Delta z_1 e^{b M_0 + c M_1}
\end{split}
\end{align}

Finally, this methodology can be applied to $\frac{dQ}{dt}_{max}$ for its relationship with $M_0$, $M_1$, and $M_2$. This takes the form of:
\begin{align}
\begin{split}
    \frac{dQ}{dt}_{max} &= Q_0(M_1, M_2) e^{-d M_0} \\
    Q_0 &= Q_1(M_2) e^{f M_1} \\
    Q_1 &= Q_2 e^{-g M_2} \\
    \frac{dQ}{dt}_{max} &= Q_2 e^{-d M_0 + f M_1 - g M_2}
\end{split}
\end{align}

\section{Discussion and Conclusions}
Minkowski functionals have shown promise in their ability to describe geometrially-influenced complex mesoscale phenomenon in porous media. Through the use of surface CRNs--a unique model of asynchronous cellular automata--to model dissolution behavior in chemical systems, the effects of Minkowski functionals on chemical behavior were extracted. Due to the unique challenges of modeling and characterizing interfacial chemical reactions, the effects of individual simulation hyperparameters were examined to understand their impact on equilibrium metrics, namely $K_{eq}$. Reaction rate scaling showed a simple log-linear relationship in dictating $K_{eq}$ behavior and the dissolution rate in the system appeared to have a direct effect on $K_eq$. This verifies previous literature that has shown discrepancies in the classical Law of Mass Action and true $K_{eq}$ values of non-well mixed systems, with these discrepancies related to energetic considerations tied directly to interphase behavior and reaction rates. Beyond the modeled chemistry influence on $K_{eq}$ valuation, unique artifacts of the surface CRN simulator must also be taken into account. Specifically, the nature of the reaction selected introduces a branching interface diffusion phenomenon even in systems of no assigned chemical diffusion. This adds an additional layer of slow manifold evolution with potential direct effects on equilibrium behavior of the system.
\newline
Ultimately, exponential relationships were found for $K_{eq}$, $\frac{dQ}{dt}_{max}$, and $\Delta \tau$ and extracted Minkowski functionals. With this linkage found and the appropriate scaling quantified, this work stands as an important step in further understanding how Minkowski functionals influence microstructural behavior.



\bibliography{bibliography}

\end{document}